\documentclass{aastex631}
\hypersetup{linkcolor=blue,citecolor=cyan,filecolor=cyan,urlcolor=blue}

\shorttitle{Properties of Turbulent Convection and Large-Scale Flows in a Rotating F-type Star}
\shortauthors{Kitiashvili et al.}

\begin{document}
\title{Properties of Turbulent Convection and Large-Scale Flows in a \\ Rotating F-type Star Revealed by 3D Realistic Radiative Hydrodynamic Simulations}

\correspondingauthor{Irina~N. Kitiashvili}
\email{irina.n.kitiashvili@nasa.gov}

\author[0000-0003-4144-2270]{Irina~N. Kitiashvili}
\affiliation{NASA Ames Research Center\\
Moffett Field, Building N-258\\
Mountain View, CA 94035, USA}

\author[0000-0003-0364-4883]{Alexander~G. Kosovichev}
\affiliation{NASA Ames Research Center\\
Moffett Field, Building N-258\\
Mountain View, CA 94035, USA}
\affiliation{New Jersey Institute of Technology\\
323 Dr Martin Luther King Jr Blvd\\
Newark, NJ 07102, USA}

\author{Alan~A. Wray}
\affiliation{NASA Ames Research Center\\
Moffett Field, Building N-258\\
Mountain View, CA 94035, USA}

\begin{abstract}
The nonlinear coupling between stellar convection and rotation is of great interest because it relates to understanding both stellar evolution and activity. We investigated the influence of rotation and the Coriolis force on the dynamics and thermodynamic structure of an F-type main-sequence star with a shallow outer convection zone. We performed a series of 3D radiative hydrodynamic simulations of a 1.47M$_{\odot}$ star for different rotation rates (periods of rotation 1 and 14 days) and with computational domains placed at latitudes of $0^\mathrm{o}$ (equator), $30^\mathrm{o}$, and $60^\mathrm{o}$. Because the star has a relatively shallow convection zone (28.5~Mm thick or about 2.81\% R$_{*}$), we model its dynamics from the upper layers of the radiative zone, the whole convection zone, and the low atmosphere. The simulation results show a weak shift of the ionization zones to the photosphere and a decrease of the stellar radius by about 29 km at the equator and about 58 km at higher latitudes in the presence of rotation with a period of 1~day. The models presented reveal the formation of radial differential rotation, meridional flows, latitude-dependent roll-like structures of convection, a tachocline, the presence of a gravity-darkening effect, and others. In this paper, we primarily discuss the properties of the outer convection zone for different rotation rates. Detailed analysis of the properties of the tachocline, the overshoot layer, and small-scale turbulence will be discussed in follow-on papers.
\end{abstract}


\keywords{Stellar rotation(1629) --- Stellar convective zones(301) --- Gravity darkening(680) --- Hydrodynamical simulations(767) ---  Radiative transfer simulations(1967)}

\section{Introduction}\label{sec:intro}

Stellar rotation is one of the fundamental characteristics that determine the properties of the large-scale flows inside stars, such as differential rotation and meridional circulation, that link to magnetic activity \citep[e.g.,][]{Noyes1984b,BoehmVitense2007,Reiners2014}. Therefore, understanding the coupling of rotation with the dynamic and thermodynamic properties of convection is critical to uncovering processes that shape subsurface structure, large-scale shearing flows, and stellar activity, and to help reveal a correlation with observed global stellar parameters \citep[e.g.,][]{Roxburgh1966,Noyes1984a,Noyes1985,Mangeney1986}. 
Because of the limited ability to gain knowledge about stellar surface and subsurface dynamics, stellar differential rotation is usually described qualitatively, relative to the Sun, as being of two types: solar-like (faster rotation at the equator and slower at the poles) or anti-solar, whereas actual variations in the latitudinal gradient of rotation may be significantly more complicated than just these two types. In this context, observations suggest that most main-sequence stars have solar-like differential rotation \citep[e.g.,][]{Reiners2006,Reinhold2013a}, and that rotation can be interpreted as anti-solar only for a few stars \citep{Reiners2003,AmmlervonEiff2012,Noraz2022a}. This disproportion in detecting solar- and anti-solar-type differential rotation raises questions about the coupling between highly stratified turbulent convection and rotation. In particular, questions arise about how to determine a specific radial and latitudinal differential rotation distribution and how rotation impacts the dynamics and convection-zone structure.

In the absence of spatial resolution in stellar observations, it is important to perform detailed theoretical and numerical studies. Investigations of stellar convection in the presence of rotation are primarily conducted at global scales using spherical shell or wedge models and at local scales using models constrained to a rectangular box. Modeling stellar dynamics on global scales naturally includes the effects of the Coriolis and centrifugal forces. However, due to a number of numerical challenges, most global models are performed using an anelastic approximation without modeling radiative transfer. Thus, they exclude the near-surface layers responsible for the radiative cooling that drives convective motions, thereby playing an important role in mixing different layers of the convection zone and transporting convective energy and momentum across a wide range of turbulent scales. On the other hand, global models allow testing different regimes of convective motions in the presence of rotation \citep[e.g.,][]{Guenther1985,Gilman1986,Miesch2000,Brun2002a,Brun2017,Kapyla2011,Kapyla2023,Guerrero2013,Beaudoin2018,Hindman2020,Warnecke2020,Hotta2022}.

It is important to note that global models tend to produce an anti-solar differential rotation in cases with a rotation rate close to solar. Increasing the rotation rate corrects this mismatch but also drives the formation of Taylor-Proudman columns, which contradicts the known distribution of the radial rotation rate in the Sun. The situation became even more intriguing after the discovery that the inclusion of even weak magnetic fields changes the differential rotation regime to solar-like \citep{Fan2014}. Further studies confirmed this result and suggested that fast flows at the equator are driven by meridional flows generated by Maxwell stresses \citep{Hotta2022}. Such a dramatic change in the properties of global flows in the presence of weak magnetic fields suggests that existing global models probably cannot correctly capture the nonlinear coupling of rotation with a wide range of turbulent scales in the convection zone as they lack accurate modeling of the turbulent viscosity and diffusion. Since turbulent diffusion is a phenomenon that cannot be quantified from observations, improvements in turbulence modeling are needed in these models. To address the issue of uncertainty in the turbulent properties of convection on global scales, \cite{Warnecke2020} imposed trial turbulent transport coefficients for a wide range of stellar rotation rates and other parameters to explore the impact on the global properties of convection and possible global dynamo mechanisms. However, comparing these results with observations is a challenging problem. Therefore, the current effort is concentrated mainly on improving global models by using a more realistic internal stellar structure, taking into account the effects of compressibility, accurate mimicking or direct simulation of radiative cooling at the photosphere, and others \citep[e.g.,][]{Guerrero2013,Nelson2018,Matilsky2020,Brun2022,Hotta2022}. 

The local approach simulates the dynamics of a small fraction of the stellar interior, including the surface and subsurface layers, and resolves the processes that drive stellar convection. This approach allows us to investigate the nonlinear dynamical coupling between rotation and turbulent radiating plasma on scales not resolved by global models. 
This approach also allows us to obtain a more realistic model by accurately modeling radiative energy transfer, chemical composition, effects of small-scale turbulence, and others. Previously, the advantage of local models to study the coupling of rotation and convection was demonstrated for simulation domains located in the convection zone at different latitudes \citep{Pulkkinen1993}. Despite a relatively simple formulation and low spatial resolution, these models could capture the change of anisotropy degree with latitude, estimate turbulent transport coefficients, and demonstrate the importance of Reynolds stresses to describe angular momentum transport. 
A similar approach was earlier applied to study the effects of the Coriolis force on the thermodynamic structure and energy transport in a local rectangular domain \citep{Brummell1996,Brummell1998}, assuming a perfect gas and for various Rossby numbers. It was found that the turbulent structure of convection remains the same for no rotation, whereas the thermodynamic mixing properties and temporal variations of the convective motions are affected by inertial motions due to rotation.
A more detailed study to understand puzzles found in global stellar models used direct numerical simulations \citep{Barekat2021} for the case of forced turbulence in unstratified plasma subject to isothermal convection with a constant sound speed. These simplifications allowed the generation of models to study convective flows at different latitudes and depths.

\begin{figure}[b]
	\begin{center}
		\includegraphics[scale=1]{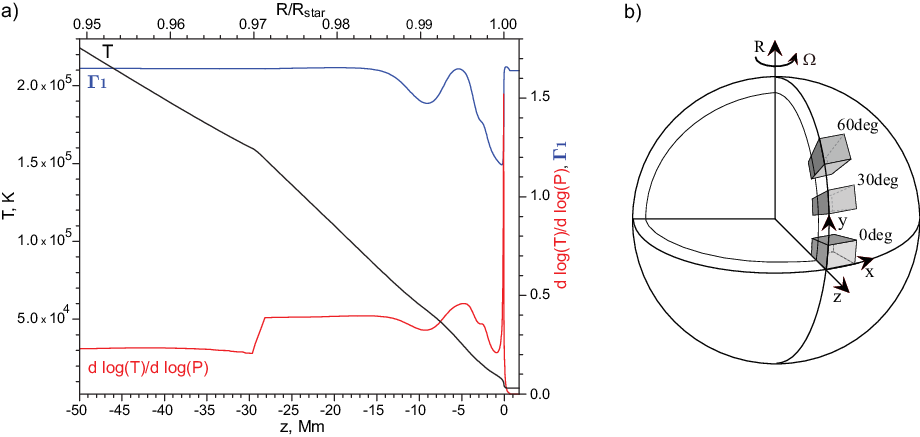}
	\end{center}
	\caption{Panel a: Radial profiles of temperature, T, temperature gradient (${\rm d log(T)/d log (P)}$), and adiabatic index, $\Gamma_1$ corresponding to the initial conditions of a 1.47M$_\odot$ main-sequence star obtained with the CESAM code. Panel b: A schematic illustration of the positions and orientations of the simulation boxes relative to the rotation axis of the star. \label{fig:scheme}}
\end{figure}

Currently, 3D realistic (`ab initio') hydrodynamic and MHD modeling have made significant progress in reproducing many observed phenomena on the Sun \citep[e.g.,][]{Nordlund2001,Cheung2007,Stein2011,Rempel2011,Rempel2018,Kitiashvili2013,Kitiashvili2023,Kitiashvili2025,Chen2023} and have been extended to other stars \citep[e.g.,][]{Beeck2012,Magic2013,Magic2014, Trampedach2013,Kitiashvili2016,Kitiashvili2023,Salhab2018,Granovsky2023}.
Recently, this approach was utilized in 3D realistic-type radiative hydrodynamic simulations of the upper layers of the solar convection zone \citep{Kitiashvili2023}. These simulations showed that the interaction of rotation and turbulent convection leads to the development of meridional flows and a fine-scale structure in the Near-Surface Shear Layer, the so-called `leptocline,' in agreement with observations at the solar photosphere and using helioseismic inferences. Interestingly, the simulations revealed weak overshoot flows at the bottom of this near-surface layer at a depth of $\sim 1.15\%~R_\odot$ (or $\sim$ 8-Mm). In contrast to the overshoot at the interface of the convection and radiative zones, the leptocline represents a boundary between more and less convectively unstable layers due to changes in hydrogen and helium ionizations \citep{Kitiashvili2023}.

In this work, we investigate the impact of rotation, particularly the Coriolis force, on the thermodynamics, structure, and dynamics of a star more massive than the Sun. Because more massive F-type stars in the main sequence tend to have shallower convection zones than the Sun, these stars cannot be studied with global models due to the strong effects of plasma compressibility, intense mixing, and the importance of radiative transfer. At the same time, the shallow convection zone allows us to include the upper layers of the radiative zone and the whole outer convection zone in the computational domain, thus allowing simulations with a high degree of realism. 

In Section~\ref{sec:setup}, we describe the numerical setup of the simulations discussed in the paper. After this, we will review the general structure of stellar convective flows (Section~\ref{sec:structure}), the properties of large-scale flows (Section~\ref{sec:global_flows}), followed by a discussion on thermodynamic structure (Section~\ref{sec:thermodyn}) and the self-development the latitude dependence of roll-like structures (Section~\ref{sec:rolls}). Finally, in Section~\ref{sec:conlusion}, we will describe the overall similarities and differences in the properties of convection zone dynamics and structure for the two imposed rotation rates at three latitudes.

\section{Computational setup} \label{sec:setup}

To investigate the influence of rotation on the structure and dynamics of stellar interiors, we performed a series of 3D hydrodynamic radiative models of a rotating spectral F-type main-sequence star of M1.47~M$_\odot$. The initial conditions were obtained with the CESAM stellar evolution code \citep{Morel1997,Morel2008}. The initial conditions used are for a star of the following properties: M=1.47~M$_\odot$, T$_{eff}$=7063~K, log(g~(cm s$^{-2}$))=4.279, L/L$_\odot$=4.731, R=1.456~R$_\odot$, [Fe/H]=0.0 for age 1~Gyr \citep{Kitiashvili2016}. The initial conditions reveal a $\sim 28.5$~Mm ($\sim 2.81$\%~R$_{*}$) thick outer convection zone. Hydrogen and helium ionization zones occupy a roughly 15~Mm thick layer of the outer convection zone (Fig.~\ref{fig:scheme}a).

The resulting 3D radiative hydrodynamic models were obtained using the StellarBox code \citep{Wray2018}, where computations are performed from first principles and take into account a realistic chemical composition and equation of state, plasma compressibility, radiative transfer, and the effects of both resolved and subgrid-scale (SGS) turbulence. The simulations were performed in Cartesian geometry. The computational models were developed for solar composition. The SGS turbulence models \citep{Smagorinsky1963,Moin1991} were implemented to describe small-scale energy dissipation and transport. In this simulation, we utilized the compressible Smagorinsly model with turbulent coefficients of $\rm C_C=C_S=0.01$, initially determined for modeling solar convection to simulate resolved granulation and intergranular flows in agreement with observations. The radiative transfer calculations are performed for four spectral bins using ray-tracing along 18 directional rays \citep{Feautrier1964} and the long-characteristics method. The wavelength-dependent opacity data was obtained from a code provided by the Opacity Project \citep{Seaton1995,Badnell2005}. The boundary conditions are periodic in the horizontal directions. The top and bottom boundary conditions are implemented by a characteristic method \citep[e.g.,][]{Sun1995}. The bottom boundary of the computational domain is open for radiation and simulates the energy input from the stellar interior. The code details, implementation, and testing are described in \citet{Wray2015,Wray2018}.

\begin{figure}[b]
	\begin{center}
		\includegraphics[scale=1.2]{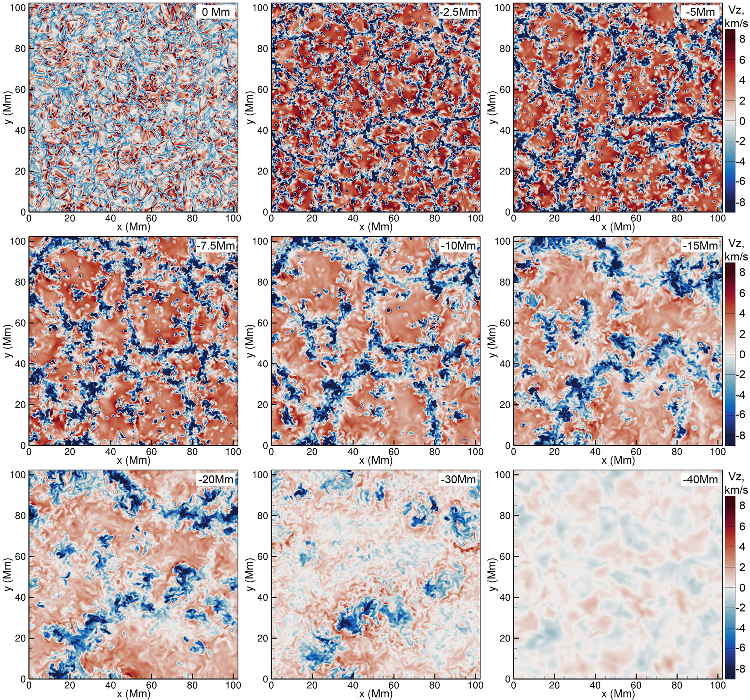}
	\end{center}
	\caption{Vertical velocity distribution at the equator in nine layers from simulations of a 1.47~M$_\odot$ star with imposed rotation for the whole domain with a period of 1 day. The horizontal slices correspond to layers throughout the convective zone and into the radiative zone, 0~Mm (R$_0 = $ R$_{*}$), 2.5, 5, 7.5, 10, 15, 20, 30, and 40~Mm below the surface. \label{fig:xy}}
\end{figure}

The computational domain of all models covers about 5\% of the stellar radius (or 50.5~Mm in depth). Thus, the models cover the stellar interior from the upper radiative zone, through the entire convection zone, and into the low atmosphere. The simulations include a 1~Mm-hight atmospheric layer. The horizontal size of the computational domain is 102.4~Mm (about $5.8^{\mathrm{o}}$) with a resolution of 100~km; the vertical resolution increases with depth from 25~km in the atmosphere to 183~km near the bottom boundary. To model the effects of rotation, we use the $f$-plane approximation, where the imposed rotational period is constant within the computational domain so that differential rotation and meridional circulation are developed naturally. The effects of surface curvature and centrifugal force are neglected. The resulting models were obtained for 1 and 14 days rotation periods at three latitudes: $0^{\mathrm{o}}$ (equator), $30^{\mathrm{o}}$, and $60^{\mathrm{o}}$ in the northern hemisphere (Fig.~\ref{fig:scheme}b), where the $x$-axis is oriented in the direction of stellar rotation in the azimuthal plane, the $y$-axis points toward the North pole in the meridional plane, and the $z$-axis points in the outward radial direction.

\begin{figure}[b]
	\begin{center}
		\includegraphics[scale=1]{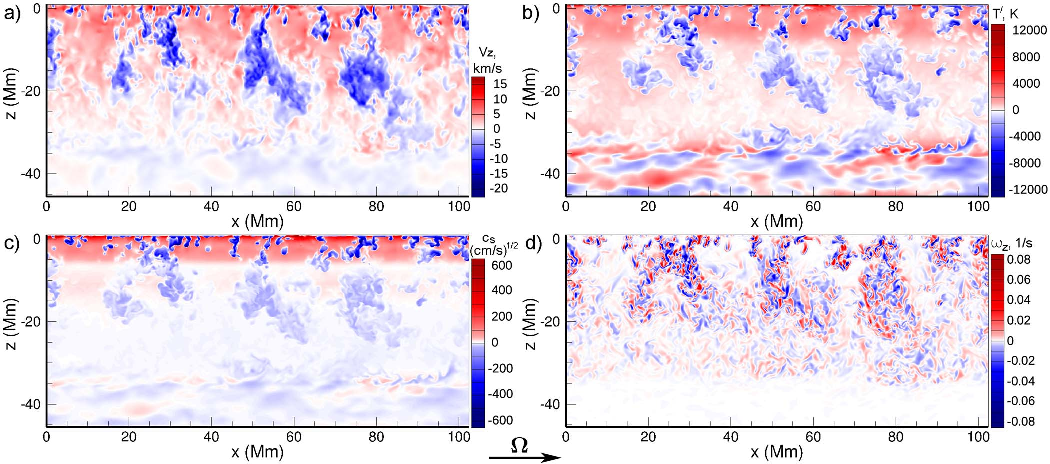}
	\end{center}
	\caption{Vertical slices through the computational domain show the distribution of a) the vertical velocity, b) temperature fluctuations, c) normalized fluctuations of the sound speed, and d) vertical vorticity in the azimuthal plane at the equator for P$_{\rm rot}=$1~day. \label{fig:VarXZ}}
\end{figure}

\section{Convection Structure of the Stellar Interior}\label{sec:structure}

Simulations of a main-sequence star of a 1.47M$_\odot$ star reveal the co-existence of several granulation scales. The larger granules, $\sim 10-12$~Mm, tend to organize in clusters with convective patterns similar to solar mesogranulation. These clusters are surrounded by smaller granules ($\sim 2$~Mm). The characteristic size of the large and small granules corresponds to the depth of the HeI and H ionization zones \citep{Kitiashvili2016}. In deeper layers of the convection zone, the scale of the convection patterns and the widths of the downflow lanes (similar to the intergranular lanes at the photosphere) gradually increase (Fig.~\ref{fig:xy}). Due to strong turbulent motions and intense mixing within the convection zone, the turbulence remains inhomogeneous. 

Similarly to the Sun, in the photosphere and subphotospheric layers the stronger helical motions are primarily distributed at the edges of granules and in the intergranular lanes. In the deeper layers, the turbulent flows become stronger inside the upflow patterns due to the intense mixing. Near the stellar tachocline, the interface between the convection and radiative zones at z $\sim -25.8$~Mm, the more intense turbulent flows surround overshooting downdrafts (Fig.~\ref{fig:xy}). 

The convective downdrafts reveal two types of downflows: shallow ones with a depth of penetration of about $10-15$Mm and high-speed ones that can penetrate the entire convection zone, reaching velocities of 20 -- 25~km/s. 
It is important to note that most downdrafts have a slightly inclined direction of rotation due to shear caused by self-formed differential rotation. However, the influence of rotation may not be exhibited for shallow downdrafts and/or high-speed downflows.
The strongest downdrafts can partially penetrate the radiative zone, initiating strong overshoot motions accompanied by vorticity generation, which also drives enhancement of turbulence along the convection-radiative zone interface (Figs.~\ref{fig:xy},~\ref{fig:VarXZ}a,d). This intense interaction of the downdrafts and the upper layers of the radiative zone causes a local compression and heating of the plasma accompanied by excitation of acoustic and internal gravity waves \citep{Kitiashvili2016}, as well as increasing variations in the sound speed (Fig.~\ref{fig:VarXZ}b,c), and initiates density fluctuations that are a source of internal gravity oscillations ($g$-modes). Thus, in the convectively stable region, the radiative zone (below 35~Mm), the remaining fluctuations represent remnants of the strong downflows that hit the tachocline. 

\section{Large-scale dynamics} \label{sec:global_flows}
The investigation of the properties of large-scale flows is crucial for understanding global magnetic activity and stellar evolution. By large scales, we mean the convective dynamics operating on scales comparable to the computational box size in the lateral directions and the thickness of the convection zone in the vertical direction.  Because the modeling is performed in a rectangular box extending over a $\sim 5.8^{\mathrm{o}} \times 5.8^{\mathrm{o}}$ area of the star and includes layers of the upper radiative zone, the whole convection zone, and photosphere, these dimensions define what we mean by ``large scales." In the presented models, we examine the influence of the Coriolis force at two rotation rates, P$_{\rm rot}=1$ day and P$_{\rm rot}=14$~days, and at three latitudes ($0^\mathrm{o}$, $30^\mathrm{o}$, and $60^\mathrm{o}$). In this section, we discuss the properties of the self-developed differential rotation relative to the imposed rotation rate, the meridional flows, and the formation of roll-like structures.

\begin{figure}[b]
	\begin{center}
		\includegraphics[scale=1.3]{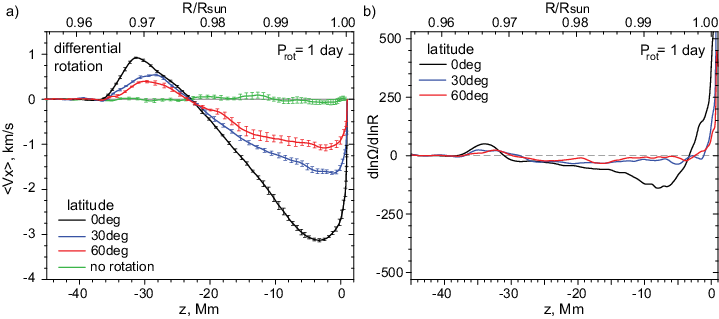}
	\end{center}
	\caption{Radial profiles of the differential rotation (panel a) and radial profiles of the gradient of rotation, $\frac{\partial\ln\Omega}{\partial\ln r}$ (panel b) for P$_{\rm rot}=1$~day at different latitudes: $60^{\mathrm{o}}$ (equator, black curves), $30^{\mathrm{o}}$ (blue), and $60^{\mathrm{o}}$ (red curves). The green curve corresponds to the non-rotating case \citep{Kitiashvili2016}. Each profile has been obtained by averaging over space and time (over one 1-hour) for the azimuthal velocity component, $V_x$. The vertical bars show the standard deviation from the mean flows. \label{fig:diff-rot1}}
\end{figure}

\subsection{Differential rotation}
To study the impact of the Coriolis force, we apply a constant rotation rate to the whole computational domain. Because of the strong stratification of the stellar plasma, it is not surprising that the mean azimuthal flow deviates from the imposed rotation rate. For faster rotation (P$_{\rm rot}=1$~day), the azimuthal component of flows, $V_x$, shows significant deceleration, up to 3.1~km/s at the equator (Fig.~\ref{fig:diff-rot1}a, black curve) and about 1~km/s at the $60^\mathrm{o}$ latitude (red curve). At the equator, the strongest deceleration occurs at depths of 3.3~Mm. For higher latitudes, the slowest flows are shifted closer to the photosphere to a depth of 2.5~Mm. In the deeper layers, the azimuthal component of flows gradually accelerates and becomes faster than the imposed rotation speed at $24-25$~Mm below the surface. The maximum rotation speed is latitude-dependent. The strongest flows correspond to the equatorial plane at a depth of 32~Mm. The maximum speed decreases with latitude, shifting the maximum rotation rate  $2-4$~Mm closer to the photosphere. Such correlation is consistent with the conservation of angular momentum. After reaching the maximum, the azimuthal flow decelerates to the imposed rotation rate. In the radiative zone (z$=-36$~Mm and deeper), the stellar interior exhibits solid-body rotation (Fig.~\ref{fig:diff-rot1}a), though small deviations occur as remnants of the dynamics in the convection zone. It is interesting to note that the gradient of rotation is nearly constant, $\partial\ln\Omega/\partial\ln r \sim 25$ for $30^\mathrm{o}$ and $\sim15-20$ for $60^\mathrm{o}$ latitudes (red and blue curves in Fig.~\ref{fig:diff-rot1}b), whereas at the equator, the gradient reaches a minimum ($\sim 140$) at $\sim 7$~Mm below the photosphere.

\begin{figure}[t]
	\begin{center}
		\includegraphics[scale=1.3]{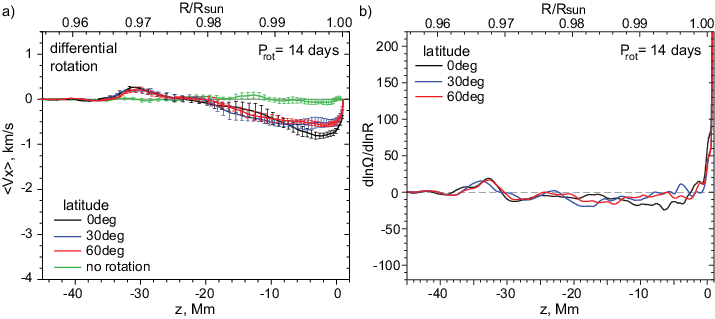}
	\end{center}
	\caption{Radial profiles of the differential rotation (panel a) and radial profiles of the gradient of rotation, $\partial\ln\Omega/\partial\ln r$ (panel b) for P$_{\rm rot}=14$~days at different latitudes: $60^{\mathrm{o}}$ (equator, black curves), $30^{\mathrm{o}}$ (blue), and $60^{\mathrm{o}}$ (red curves). The green curve corresponds to the non-rotating case \citep{Kitiashvili2016}. Each profile has been obtained by averaging over space and over one hour in time for the azimuthal velocity component, $V_x$. The vertical bars show the standard deviation from the mean flows. 
	\label{fig:diff-rot14}}
\end{figure}

In the case of slower rotation (P$_{\rm rot}=14$~days, Fig.~\ref{fig:diff-rot14}), the dynamical structure of the azimuthal flows is qualitatively similar with substantially weaker latitudinal dependence. In particular, the azimuthal flow at the equator decelerates faster near the stellar surface, while flows at higher latitudes have similar variations. Below $\sim 1$\% R$_*$ (a depth of $\sim -10$~Mm ), the general properties of azimuthal flows do not exhibit latitudinal dependence. The rotation gradient is nearly constant, $\partial\ln\Omega/\partial\ln r \sim 10$, through most of the convection zone at the equator. In the case of higher latitudes, the logarithmic rotational gradient is close to zero near the photosphere and saturates near 10 below $z$ = --10~Mm depth.

\subsection{Meridional flows}
\begin{figure}[t]
	\begin{center}
		\includegraphics[scale=1.3]{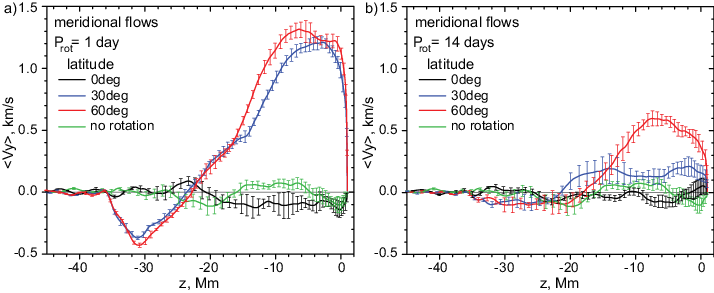}
	\end{center}
	\caption{Radial profiles of the meridional flows for P$_{\rm rot}=1$~day (panel a) and P$_{\rm rot}=14$~days (panel b) at three latitudes: $0^\mathrm{o}$ (equator, black curves), a $30^\mathrm{o}$ (blue), and $60^\mathrm{o}$ (red curves). The green curve corresponds to the non-rotating case \citep{Kitiashvili2016}. Each profile has been obtained by averaging over space and time for the velocity component perpendicular to the stellar rotation, $\rm{V_y}$. The vertical bars show the standard deviation from the mean flows. \label{fig:merid}}
\end{figure}

In the presence of rotation, all models exhibit self-development of meridional flows at medium and high latitudes but not at the equator (Fig.~\ref{fig:merid}). The meridional flows are strongest at $60^\mathrm{o}$ latitude. In the upper layers of the convection zone, the meridional flows are northward with a maximum of about 1.35~km/s at 6~Mm below the photosphere for $60^\mathrm{o}$ latitude and P$_{\rm rot}=1$~day (red curve, Fig.~\ref{fig:merid}a) and gradually decrease with depth. It is interesting to note that, in the upper convection zone, the meridional flows exhibit finer-scale structure, similar to the `leptocline' discovered in solar rotation models \citep{Kitiashvili2023} and observations \citep[e.g.,][]{Deubner1979,Rozelot2009,Komm2021}. Similarly to the leptocline in the solar convection zone  \citep{Kitiashvili2023}, the fine-scale structure co-locates with the hydrogen and helium-I ionization zones. 
At $30^\mathrm{o}$ latitude (blue curve), the meridional flows are noticeably slower ($\sim 1.2$~km/s ) and qualitatively similar, with maxima of northward flow at $\sim 2 - 4$Mm below the photosphere (blue curve, Fig.~\ref{fig:merid}a). At a depth of 14 -- 15~Mm below the photosphere, the sudden change in the mean meridional flow profile corresponds to the bottom of the HeII ionization zone. At the bottom part of the convection zone, at about --22~Mm, the meridional flows start to change direction toward the equator and reach a maximum speed at a depth of 31~Mm. As expected, there is no presence of meridional flows at the equator. The meridional component of the flows ($\rm{V_y}$) shows properties of the non-rotating plasma (green curve, Fig.~\ref{fig:merid}a).

In the case of the slower rotation (P$_{\rm rot}=14$~days, Fig.~\ref{fig:merid}b), the difference between the $30^\mathrm{o}$ and $60^\mathrm{o}$ latitudes is more significant in the upper layers of the convection zone, where the maximum speed of the meridional flows reaches about 0.65~km/s at $60^\mathrm{o}$ latitude (red curve) and 0.2~km/s at $30^\mathrm{o}$ latitude (blue). In the upper layers of the convection zone, the gradient of the meridional flows at $60^\mathrm{o}$ latitude (red curve) is similar to the faster rotation case. In particular, the meridional flow speed quickly rises to about 0.5~km/s at around a depth of 4~Mm and continues a slower rise in deeper layers that correspond to the bottom of the leptocline.
For the medium latitude ($30^\mathrm{o}$, blue curve), the northward meridional flow is about 0.2~km/s in the bulk of the convection zone. The return flows are significantly weaker (about --0.1~km/s) in the slow rotation case and do not show a visible latitudinal dependence in their speed. A latitudinal dependence is noticeable in the thickness of the layer occupied by the reverse flows. In particular, at $60^\mathrm{o}$ latitude, the return flows start below 18~Mm (approximately the middle of the convection zone), whereas at $30^\mathrm{o}$ latitude, this depth is about 21 -- 22~Mm.

\begin{figure}[b]
	\begin{center}
		\includegraphics[scale=0.9]{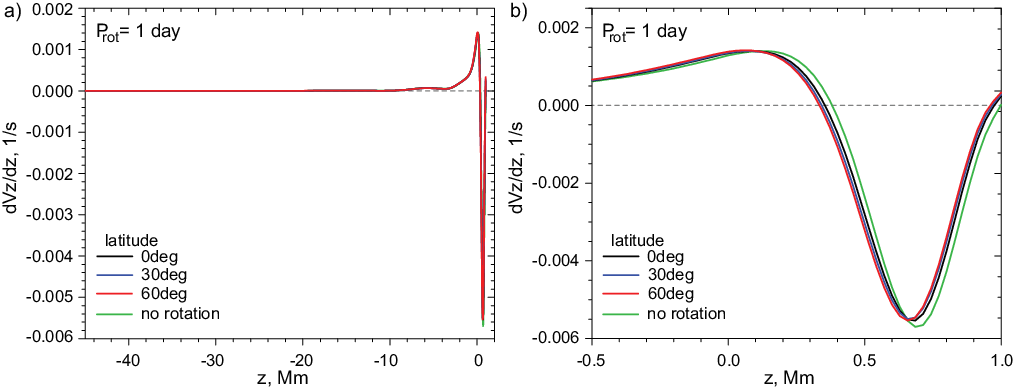}
	\end{center}
	\caption{The radial gradient of the vertical velocity over the computational domain (panel a) and the near-surface layers (panel b) for P$_{\rm rot}=1$~day at different latitudes.
 \label{fig:dVzdz}}
\end{figure}

\subsection{Radial flows}
Generally, the radial flows in this star are significantly stronger than solar convection flows. For instance, the downflows can reach 20 -- 25 km/s, penetrate through the whole stellar convection zone, and overshoot into a convectively stable radiative zone, exciting acoustic and internal gravity waves and forming an overshoot layer (Fig.~\ref{fig:VarXZ}), whereas in quiet-Sun regions, downdrafts occur mostly in a 2~Mm-thick subsurface layer \citep[e.g.,][]{Kitiashvili2025}. Because of the high speed of the downflows in the modeled F star, they are not very sensitive to the background rotation, though some weaker downflows show a gradual inclination in the direction of the rotation (Fig.~\ref{fig:VarXZ}a). Because of the strong overshooting at the interface between the convection and radiative zones (tachocline), the bottom of the convection zone becomes more turbulent and experiences intense mixing. The radial gradient of the vertical velocity (Fig.~\ref{fig:dVzdz}a) exhibits the most significant variations in the upper layers of the convection zone and quickly decreases with depth. Effects of rotation are noticeable only near the photosphere (Fig.~\ref{fig:dVzdz}b), where the imposed rotation causes a decrease in the radial velocity gradient due to suppression of the turbulent motions by rotation, an effect that leads to a downward shift of the photosphere by about 40~km for the 1-day period of rotation. The latitudinal dependence of the velocity shows a weak gradual displacement at the equator and is stronger for the higher latitudes. It also indicates a slow stellar radius change with latitude.

\begin{figure}[b]
	\begin{center}
		\includegraphics[scale=1.]{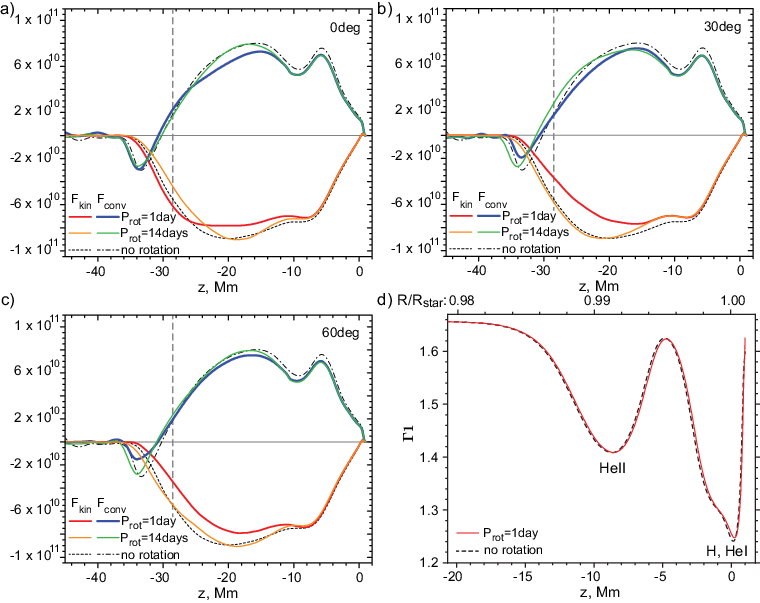}
	\end{center}
	\caption{The radial distribution of the convective energy flux, $\rm F_{conv}$, and the kinetic energy flux, $\rm F_{kin}$, associated with convection at different latitudes: a) $0^\mathrm{o}$ (equator), b) $30^\mathrm{o}$, and c) $60^\mathrm{o}$. Darker thick curves show the distribution of fluxes for P$_{\rm rot}=1$~day; the lighter thin curves correspond to P$_{\rm rot}=14$~days. Panel d) shows the adiabatic exponent for the cases with and without rotation. Dashed black curves correspond to the non-rotating case. Vertical dashed lines indicate the bottom of the convection zone. 
	\label{fig:EF}}
\end{figure}

\section{Thermodynamic properties of the stellar interiors} \label{sec:thermodyn}
Changes in the thermodynamic properties of the convection zone lead to changes in differential rotation and meridional flows. For instance, a correlation has been shown between the adiabatic exponents (associated with H, He~I, and He~II ionization zones) with the properties of the convection scales \citep{Kitiashvili2016,Kitiashvili2023}. Furthermore, the presence of rotation causes slight changes in the radial distribution of the adiabatic index (red curve; Fig.~\ref{fig:EF}d), which manifests as a slight decrease in the thickness of the H and He~I ionization zones and a shift of the He~II ionization zone toward the photosphere. 
To investigate the influence of stellar rotation on the energetics of the convection zone, we analyze properties of the convective energy flux and the kinetic energy flux associated with convection, defined following \citet{Nordlund2001} as
\begin{eqnarray}
    {\rm F_{conv}=\left\langle \left( \rho E_{i}+ P \right) \breve{V}_z\right\rangle,}\\
	{\rm F_{kin}=\left\langle \left( \frac{1}{2}\rho V^2\right) \breve{V}_z\right\rangle,}
\end{eqnarray}
where $\left\langle ... \right\rangle$ is the horizontal averaging, ${\rm \breve{V}_z=V_z-\bar{V}_z}$, ${\rm \bar{V}_z=\left\langle \rho V_z\right\rangle/\left\langle \rho\right\rangle}$, $\rho$ is density, ${\rm E_{i}}$ is the internal energy per unit mass, and P is the gas pressure.

The simulation results show that the presence of rotation causes a decrease in the convective and kinetic energy fluxes in the subsurface layers (Fig.~\ref{fig:EF}) in comparison with the case without rotation (dashed and dash-doted curves). The impact of rotation on the energy flux distribution depends on the latitude and the rotation rate. In particular, in the case of the faster rotation (P$_{\rm rot}=1$~day), the convective energy flux is lower through most of the convection zone (blue curves, Fig.~\ref{fig:EF}). At the equator, the convective energy flux is higher compared to the case without rotation near the bottom of the convection zone. In contrast, at $60^\mathrm{o}$ latitude, the convective energy flux shows a gradual convergence with the non-rotation case with weaker flux variation in the overshoot region. 
Interestingly, the kinetic energy flux shows a more significant latitudinal dependence. For instance, at the equator, the kinetic energy flux (red curves, Fig.~\ref{fig:EF}a) is primarily suppressed in the middle of the convection zone, whereas, at the higher latitudes, it is noticeably weaker from the middle of the He~II ionization zone (about 8 Mm depth) and down through the whole convection zone and the overshoot layer.

\begin{figure}[t]
	\begin{center}
		\includegraphics[scale=1]{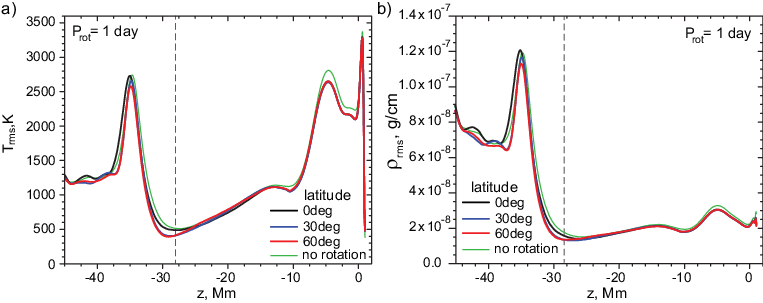}
	\end{center}
	\caption{Radial profiles of {\it rms} temperature (panel a) and density  (b) fluctuations at three latitudes: $0^\mathrm{o}$ (equator, black curves), a $30^\mathrm{o}$ (blue), and $60^\mathrm{o}$ (red curves) for the star with P$_{\rm rot}=1$~day. The green curves on panels a) and c) correspond to the stellar model without rotation. Vertical dashed lines indicate the bottom of the convection zone. \label{fig:rmsTRho}}
\end{figure}

For the lower rotation rate (P$_{\rm rot}=14$~days), the impact of rotation on the kinetic and convective energy fluxes distribution is weaker (orange and green curves, Fig.~\ref{fig:EF}). For instance, at the equator, the convective energy flux (green curves) deviates from the case without rotation (dash-dot curves) up to 16~Mm depth; this deviation disappears deeper in the convection zone. Interestingly, at $30^\mathrm{o}$ latitude, the convective energy flux is stronger near the bottom of the convection zone (below 22~Mm), which causes a noticeable displacement of the flux into deeper layers of the overshoot region. A similar weaker enhancement of the energy is present at $60^\mathrm{o}$ latitude.
The radial distribution of the kinetic energy flux associated with convection shows a slight decrease in the He~II ionization zone, becomes in agreement below He~II, and weaker at the bottom of the convection zone and the overshoot layer (Fig.~\ref{fig:EF}a). At higher latitudes, the kinetic energy flux is more in agreement with the non-rotation case for higher latitudes: below 20~Mm for $30^\mathrm{o}$ latitude, and below 13~Mm for $60^\mathrm{o}$. The greatest suppression of the energy flux occurs in the He~II ionization zone at $30^\mathrm{o}$ latitude.

Fluctuations in the temperature and density are affected by rotation mostly in the H, He~I, and He~II ionization zones and in the overshoot layer (Fig.~\ref{fig:rmsTRho}) and do not show a significant dependence on latitude. Near the bottom of the convection zone, $rms$ fluctuations are weaker compared to the no-rotation case (Figs~\ref{fig:diff-rot1},~\ref{fig:diff-rot14}).

\begin{figure}[t]
	\begin{center}
 		\includegraphics[scale=0.8]{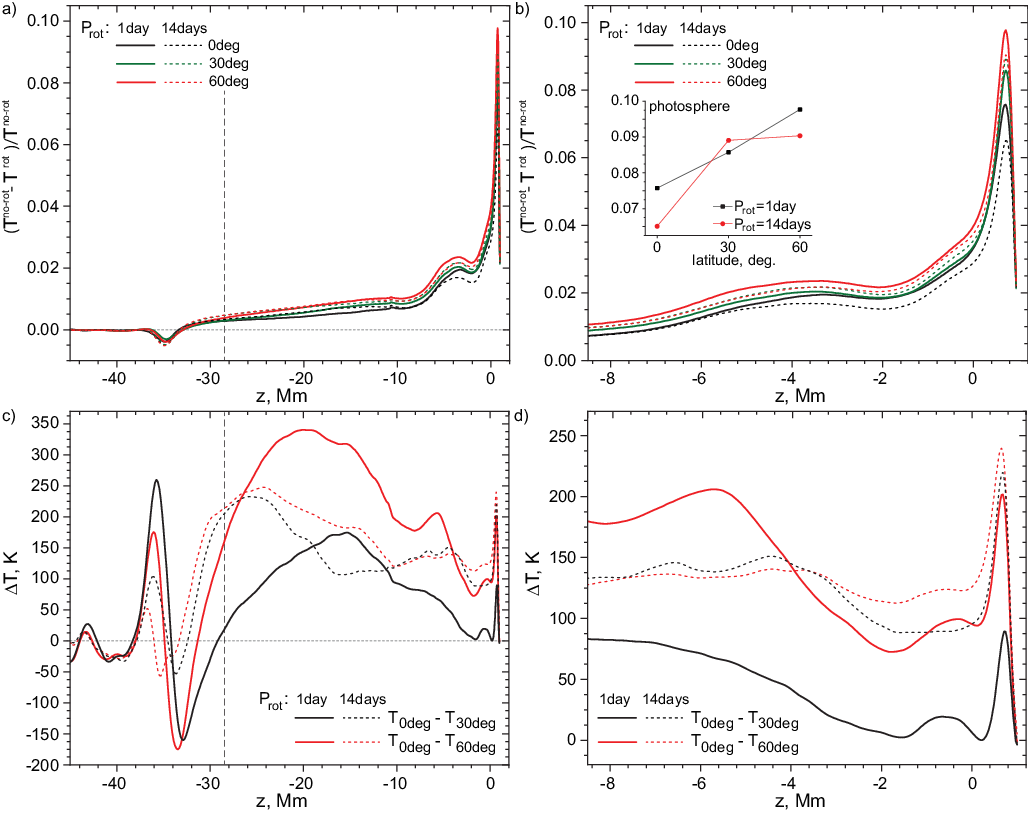}
	\end{center}
	\caption{The mean temperature relative to the non-rotating case for the whole computational domain (panel a) and magnified profiles for the upper layers of the convection zone (panel b). The inset plot in panel a) shows the relative temperature variation as a function of latitude at the photosphere.
    Panel c) shows the temperature deviations from the equator for $30^\mathrm{o}$ (black curves) and $60^\mathrm{o}$ (red curves). The magnified temperature deviations near the photosphere are shown in panel d). 
The solid curves correspond to models with P$_{\rm rot}=1$~day, and dashed curves for P$_{\rm rot}=14$~days. Vertical dashed lines indicate the bottom of the convection zone.
\label{fig:dTrot1-14}}
\end{figure}

To evaluate the influence of rotation on the global properties of convection, we consider variations of the mean temperature profile relative to the case without rotation (Fig.~\ref{fig:dTrot1-14}a, b). In general, imposed rotation causes a decrease in the mean temperature throughout the convection zone and the upper layers of the tachocline and a very small temperature increase near the bottom of the overshoot region. 
There is no change in the mean temperature in the radiative zone below 38~Mm.
The temperature decrease is greatest for faster rotation at the photosphere (Fig.~\ref{fig:dTrot1-14}a, b).
For the faster rotation case, at the photosphere the temperature deviations from the no-rotation case increase almost linearly with latitude, from about 7.5\% at the equator to about 9.75\% at $60^\mathrm{o}$ latitude. 
For the slower rotation case (P$_{\rm rot}=14$~days), the temperature deviation is about 6.5\% at the equator and increases to about 9\% at higher latitudes (see the inset plot in Fig.~\ref{fig:dTrot1-14}b).
The strong deviation of the mean temperature from the non-rotating case (up to 10\% at the equator for P$_{\rm rot}=1$~day) decreases downward through the convection zone and into the upper overshoot layer. At the bottom of the overshoot layer, the temperature of the rotating plasma is higher by about 0.05\%.

Comparison of the temperature differences relative to the equator (Fig.~\ref{fig:dTrot1-14}c, d) for P$_{\rm rot}=1$~day shows a significant temperature decrease with latitude from the photosphere through most of the convection zone and shows strong relative temperature variations in the overshoot layer. In particular, the simulation results show that the temperature decreases at the photosphere by about 90~K at $30^\mathrm{o}$ and 200~K at $60^\mathrm{o}$ latitude for P$_{\rm rot}=1$~day. For the slower rotation,  P$_{\rm rot}=14$~days, the convection zone also exhibits lower temperatures at higher latitudes, with differences at the photosphere of 220~K and 240~K respectively and reaching 235 -- 250~K at a depth of about 24 -- 26~Mm below the photosphere. In the overshoot layer, the temperature difference varies from a temperature increase in the upper layer of the overshoot region to a significant decrease near its bottom. 

\begin{figure}[t]
	\begin{center}
		\includegraphics[scale=0.2]{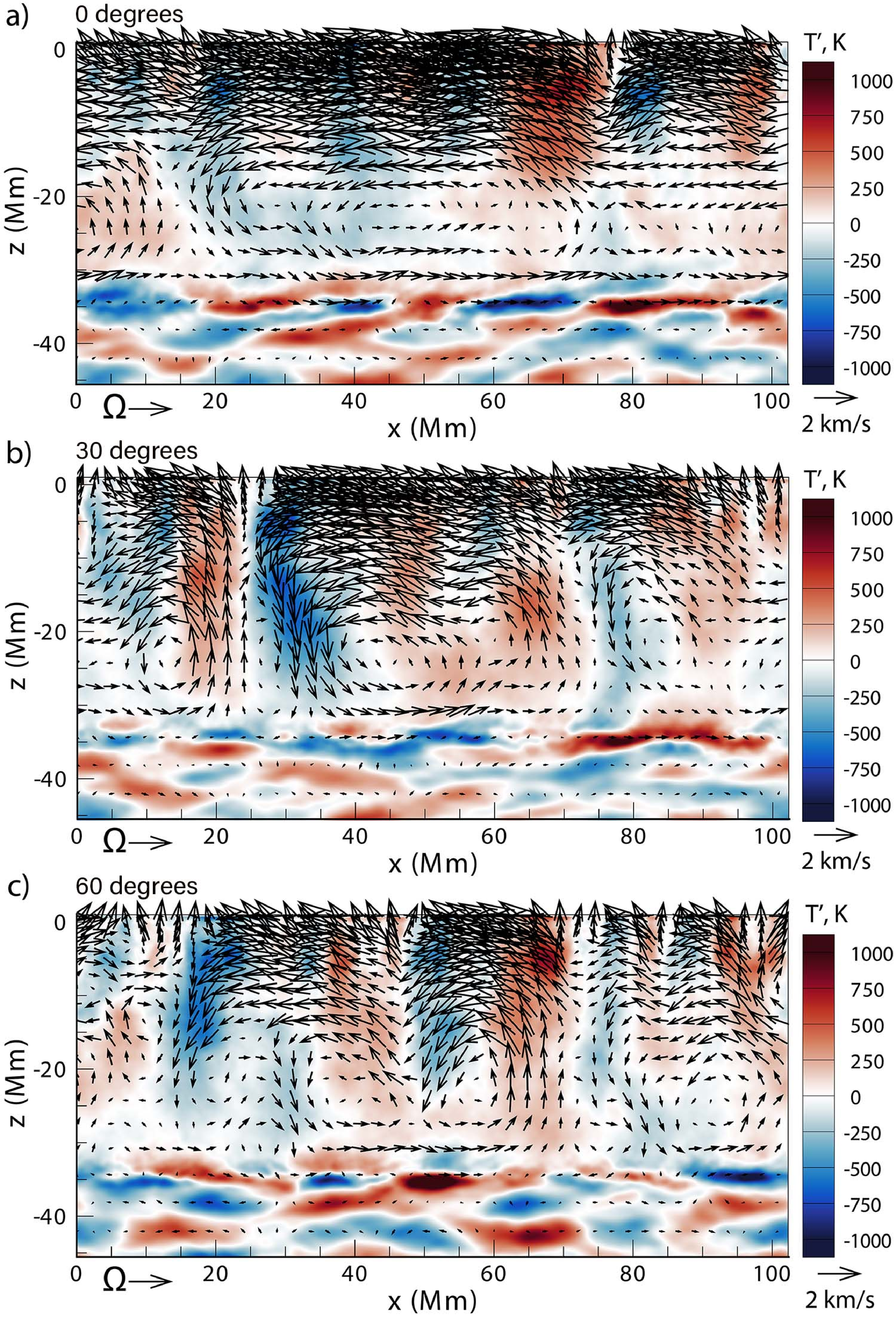}
	\end{center}
	\caption{Roll-like convective patterns for P$_{\rm rot}=1$~day at three latitudes: a) $0^\mathrm{o}$, b) $30^\mathrm{o}$, and c) $60^\mathrm{o}$. The flows are indicated by black arrows. The background images show temperature fluctuations. Flows and temperature fluctuations are averaged over the meridional plane (y-axis) to reduce noise from small-scale turbulence.
	\label{fig:rolls}}
\end{figure}

\section{Roll-like convective patterns} \label{sec:rolls}

Another view of the dynamics of the convection zone is shown in the formation of coherent structures due to interactions among the Coriolis force, turbulence, and differential rotation.
As discussed previously (section~\ref{sec:global_flows}), the differential rotation in the models has flows that are slower than the imposed angular velocity in the convection zone and faster at the bottom parts of the convection zone and in the overshoot layer (Figs~\ref{fig:diff-rot1}a,~\ref{fig:diff-rot14}a). These shearing flows cause the formation of roll-like convective structures (Fig.~\ref{fig:rolls}) that occupy the whole convection zone at the equator and become more compact at higher latitudes. These coherent structures cannot be seen in a single vertical snapshot but are visible after averaging the flow velocities and temperature perturbations along the meridional plane (Fig.~\ref{fig:VarXZ}). 

\begin{figure}[t]
	\begin{center}
		\includegraphics[scale=0.9]{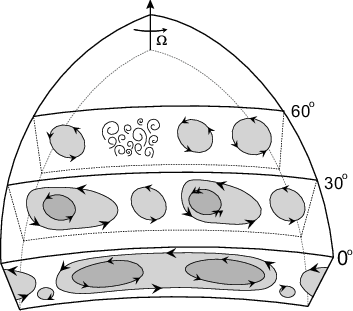}
	\end{center}
	\caption{Illustration of the roll-like structures in the convection zone at different latitudes.	\label{fig:rolls-cartoon}}
\end{figure}

At the equator, the roll-like structures are stable (relative to other latitudes), propagate in a direction opposite to the stellar rotation, and extend up to 60 -- 80~Mm in azimuth (Fig.~\ref{fig:rolls}a). It is important to note that, in the simulation, the length of the structures in the azimuthal direction can be underestimated because of the computational box size. 
Like convective motions on small scales, the upward motions are accompanied by higher temperatures and downflows are associated with lower temperatures. Inside and between these stretched large rolls, circular motions with shorter lifetimes sometimes form. The in-between small-scale rolls are typically located at the bottom of the convection zone (Fig.~\ref{fig:rolls-cartoon}).

At $30^\mathrm{o}$ latitude (Fig.~\ref{fig:rolls}b), the large roll-like structures are still present and propagating in a direction opposite to the stellar rotation. However, unlike the flow structures in the equatorial region, they are shorter in the azimuthal direction, about 40~Mm. Typically, only one additional circular substructure is formed inside these large rolls. Small-scale rolls (10 -- 15~Mm wide), located between the larger ones, are more prominent and can occupy a comparable range of layers in the convection zone. At $60^\mathrm{o}$ latitude (Fig.~\ref{fig:rolls}c), only small-scale roll-like structures with a size of 10-20~Mm are present. Sometimes, they are located next to each other, and other times they are separated by small-scale turbulent motions without definite structure.

Thus, the large-scale dynamical convection structure can be represented as roll-like patterns propagating in a direction opposite to the stellar rotation with gradual changes in the scales and complexity of these rolls with latitude (Fig.~\ref{fig:rolls-cartoon}). Small-scale circulation between the large-scale rolls can be identified. These circular motions primarily occur near the tachocline layer. At medium latitudes, the large-scale rolls are more compact in the azimuthal direction and can contain one circular substructure. The convective patterns between them have comparable scales. At the high latitude  ($60^\mathrm{o}$), the large-scale convection consists of convective rolls of smaller size that do not show finer structure. Unlike lower latitude dynamics, these rolls are distributed inhomogeneously. Turbulent regions between these patterns sometimes do not show a specific structuring and can occupy an area comparable with the rolls at this latitude (Fig.~\ref{fig:rolls-cartoon}). 

\begin{figure}[t]
	\begin{center}
		\includegraphics[scale=0.9]{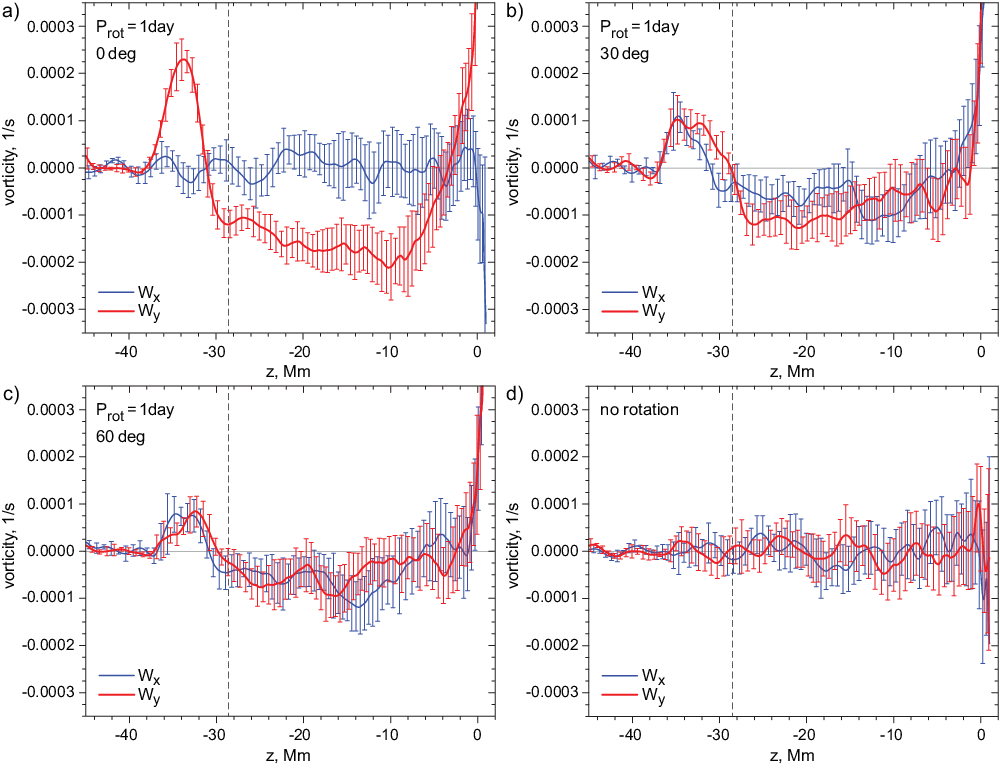}
	\end{center}
	\caption{Meridional (W$_{\rm x}$, blue curves) and azimuthal (W$_{\rm y}$, red curves) components of vorticity as a function of depth for  P$_{\rm rot}=1$~day for different latitudes: $0^\mathrm{o}$ (panel a), a $30^\mathrm{o}$ (panel b), and $60^\mathrm{o}$ (panel c), and the case without rotation (panel d). The vertical blue and red bars show the corresponding standard deviation. Vertical dashed lines indicate the bottom of the convection zone. \label{fig:vorticityXY1}}
\end{figure}

To investigate links between large-scale circular flows and small-scale helical motions, we consider the distribution of the azimuthal (${\rm W_y}$) and meridional (${\rm W_x}$) vorticity components for  P$_{\rm rot}=1$~day. At the equator (Fig.~\ref{fig:vorticityXY1}a), the azimuthal component of vorticity (red curves) is negative in most of the convective zone (except the near-photosphere layers). The mean azimuthal vorticity at $30^\mathrm{o}$ latitude remains negative. In contrast to the equatorial region, it increases with depth and reaches a maximum near the bottom of the convection zone (Fig.~\ref{fig:vorticityXY1}b). In the case of $60^\mathrm{o}$ latitude (panel c), the vorticity distribution is nearly uniform throughout the convection zone.
As expected, the meridional component of vorticity (blue curves) does not show a preferential direction in the helical motions at the equator. For higher latitudes, the predominant negative values correspond to northward meridional flows, and the vorticity sign reversal reflects the returning flows in the overshot region. 

\begin{figure}[t]
	\begin{center}
		\includegraphics[scale=0.9]{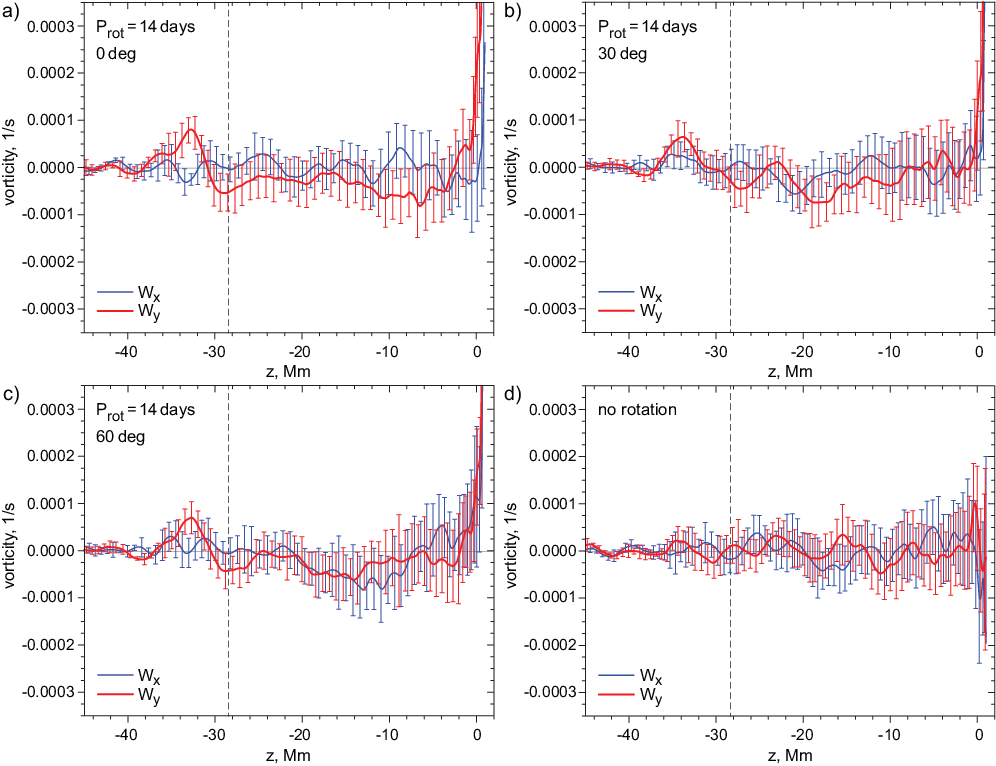}
	\end{center}
	\caption{Meridional (W$_x$, blue curves) and azimuthal components  (W$_y$, red curves) of vorticity as a function of depth for  P$_{\rm rot}=14$~days for different latitudes: $0^\mathrm{o}$ (panels a), a $30^\mathrm{o}$ (panel b), and $60^\mathrm{o}$ (panel c), and the case without rotation  (panel d). The vertical blue and red bars show the corresponding standard deviation. Vertical dashed lines indicate the bottom of the convection zone.\label{fig:vorticityXY14}}
\end{figure}

For the slower rotation (Fig.~\ref{fig:vorticityXY14}), the mean azimuthal component of vorticity  (red curves, $W_y$) remains negative in the convection zone and varies around $-5 \times 10^{-5}$~s$^{-1}$ at the equator and $60^\mathrm{o}$ latitude. In the middle latitude, $30^\mathrm{o}$, case, the vorticity variations are more significant and show preferentially counterclockwise vortical motions in the convection zone below 10~Mm. The meridional component of vorticity does not show a preferential direction of the vortical motions at the equator and at $30^\mathrm{o}$ latitude. At the higher latitude, $60^\mathrm{o}$, there is a predominance of counterclockwise motions that are strongest in the middle layers of the convection zone.

\begin{table}[th!]
    \centering
    \begin{tabular}{lll}
    \hline
    \hline
         & \bf{P$_{\rm{\bf rot}}$=1~day} & \bf{P$_{\rm{\bf rot}}$=14~days}\\
    \hline
      \bf{Differential} & 1) The retrograde flows decrease for higher & 1) The retrograde flows have a weak latitudinal \\ 
      \bf{rotation}     &    latitudes.                               & dependence.   \\
                   &  2) At the equator, the gradient of rotation is 140s$^{-1}$   &  2) At the equator, the rotation gradient is about  \\ 
                   &     at a depth of about 8~Mm. It gradually decreases       & 20 -- 25~s$^{-1}$ for the upper layers of the convection   \\
                   &     to about 25~s$^{-1}$ at the bottom of the convection   & zone and about 10~s$^{-1}$ below 10~Mm.   \\
                   &     zone. At the $30^\mathrm{o}$ and $60^\mathrm{o}$ latitudes, the rotation   & At the higher latitudes, the rotation gradient  \\
                   &     gradient is about 25~s$^{-1}$ and remains about    &  is about 10~s$^{-1}$.  \\
                   &     the same through the whole convection zone.  & \\
      \hline
      \bf{Meridional} & 1)  The maximum of the northward flows is about &  1) At $30^\mathrm{o}$ latitude, the flows have speeds\\     
      \bf{flows}      & 1.2~km/s  at a depth of about 2 -- 4~Mm for $30^\mathrm{o}$ and  & of about 0.2~km/s up to 20~Mm depth. At $60^\mathrm{o}$, \\
                      & 1.35~km/s at a depth of about 7~Mm for $60^\mathrm{o}$.      & flows reach 0.65 km/s at a depth of about 8~Mm.    \\
                      & 2) The speed of the return flows increases in the                & 2) Return flows don't have latitudinal  \\
                      &  overshoot layer. Return flows reach about                   &   dependence. The mean flows do not exhibit    \\
                      &   --0.35~km/s at $30^\mathrm{o}$ and about --0.45~km/s at    &  significant variations and have a speed of \\
                      & $60^\mathrm{o}$ latitude with maximum at depth of 31~Mm.     &  about --0.1~km/s.  \\
                      & 3) Correlation with ionization zones: A flow dip at          &  3) Correlation with ionization zones: \\   
                      &  a depth of $\sim$ 14Mm for $30^\mathrm{o}$ latitude corresponds   & A flow dip at a depth of about 4Mm at $60^\mathrm{o}$ \\   
                      & to the bottom of the HeII ionization zone. Flow            & latitude correlates with the bottom of the H \\   
                      & dips at depths of about 4Mm and 16Mm at $60^\mathrm{o}$      & and HeI ionization zones. No correlation with   \\    
                      & latitude correlate with the bottom of                       & ionization zones for the $30^\mathrm{o}$ latitude.\\
                      &  the H and HeI, and HeII ionization zones.                   & \\
      \hline
     \bf{Thermodynamical} & 1a) The mean temperature at the photosphere is           & 1a) Compared to the case without rotation, the  \\
     \bf{properties}      &  lower compared to the case without rotation by          &  mean temperature at the photosphere is lower   \\
                          &  7.5\% for $0^\mathrm{o}$, 8.5\% for $30^\mathrm{o}$, and 9.8\% for $60^\mathrm{o}$. & by 6.5\% for $0^\mathrm{o}$, 8.9\% for $30^\mathrm{o}$, and 9\% for $60^\mathrm{o}$.\\
                          & 1b) Relative to the equator, the temperature decreases     & 1b) Relative to the equator, temperature  \\
                       & by 90K at $30^\mathrm{o}$ and 202K at $60^\mathrm{o}$ at the photosphere. & decreases by 220~K at $30^\mathrm{o}$ and by 240~K at $60^\mathrm{o}$ \\  
                          & 2a) The radial distribution of the azimuthal             &  at the photosphere. \\
                          & vorticity component is strongest at the equator.         & 2a) The azimuthal component of vorticity has a \\
                          & It reaches a maximum at a depth of 10~Mm and             & predominantly negative sign and is near constant  \\
                          & decreases in the deeper layers of the convection         & through the convection zone.  \\
                          & zone. At higher latitudes, the azimuthal                 & Variations of the azimuthal vorticity   \\
                          & component of vorticity is weaker and distributed         &  component decrease with latitude in the  \\
                          & more uniformly through the convection zone.              & overshoot region. \\
                          & Variations of the azimuthal vorticity component          & 2b) The meridional component of vorticity in   \\
                          & decrease with latitude in the overshoot region.          & the convection zone has values about the same   \\
                          & 2b) The meridional component of vorticity has            & as the azimuthal vorticity component. At $60^\mathrm{o}$ \\
                          & a negative sign in the convection zone and               &  latitude, the vorticity component sign is   \\
                          &  values about the same as the azimuthal vorticity        &  predominantly negative. The meridional vorticity \\
                          &  component. At the overshoot region, vorticity           & component shows no dependence on  \\
                          &  variations are higher at $30^\mathrm{o}$ than $60^\mathrm{o}$ latitude. & latitude in the overshoot region.\\
    \hline
    \end{tabular}
    \caption{Primary differences in the convection zone's thermodynamic properties and dynamics associated with different rotation rates and latitudes.}
    \label{tab:Prot-diff}
\end{table}

\section{Discussion and Conclusions}\label{sec:conlusion}

This paper presents an analysis of the influence of rotation on the turbulent dynamics of the outer convection zone of a main-sequence F-type star of 1.47M$_\odot$ based on 3D `ab initio' radiative hydrodynamic simulations. The numerical models use a realistic stellar interior structure (used as the initial conditions), account for its chemical composition, use a realistic equation of state, account for small-scale turbulence, apply time-dependent radiative transfer, and include other physically realistic effects. To study the effects of rotation, we generated six models in a local Cartesian domain, covering the stellar convection zone from the surface to the radiative zone, with periods of rotation of 1~day and 14~days at three latitudes: $0^\mathrm{o}$ (equator), a $30^\mathrm{o}$, and $60^\mathrm{o}$ using the $f$-plane approximation for the Coriolis force (Fig.~\ref{fig:scheme}). 

The resulting models reveal the development of differential rotation and meridional flows. The differential rotation exhibits slower mean flows near the photosphere and its acceleration with the depth, which exceeds the imposed one in the overshoot layer. For the faster rotation (P$_{\rm{rot}}=1$~day), the mean flow deceleration at the near-surface layers and acceleration in the overshoot region is strongest at the equator (Fig.~\ref{fig:diff-rot1}). The impact of the stellar rotation decreases with latitude. In contrast, for slower rotation (P$_{\rm{rot}}=14$~days; Table~\ref{tab:Prot-diff}), the differential rotation at higher latitudes is distinct from the equator only at the upper layers of the convection zone (Fig.~\ref{fig:diff-rot14}). 

The resulting radial gradient of azimuthal velocities creates a shear flow that causes the formation of the roll-like structures, which may resemble Taylor-Proudman or Busse-like columns at the equator (Fig.~\ref{fig:rolls}). The structures are stretched in the azimuthal plane and can contain evolving substructures with circular motions. These rolls become more compact at $30^\mathrm{o}$ latitude and do not contain more than one substructure. At the higher latitude ($60^\mathrm{o}$), there are no large-scale structures; only small-scale circulations can be identified. Sometimes, these cyclonic motions dissolve at high latitudes. The general picture of the large-scale structuring in the stellar convection with latitude is shown in the cartoon (Fig.~\ref{fig:rolls-cartoon}), which illustrates a gradual change in the large-scale organization of dynamical structures with latitude. 
These large-scale roll-like structures display retrograde motions that explain the resulting properties of the differential rotation profiles (Figs.~\ref{fig:diff-rot1} and~\ref{fig:diff-rot14}). At the same time, because these circular motions occupy a wide range of depth, the roll-like structuring and dynamics affect the statistical properties of the flow vorticity. In particular, the mean distribution of the azimuthal vorticity component (red curves in Figs.~\ref{fig:vorticityXY1} and~\ref{fig:vorticityXY14}) shows a predominance of negative values that reflects a coupling between the stellar turbulence and rotation and is tightly linked to retrograde motions in the upper layers of the convection zone and weaker prograde motions at the bottom. At the higher latitudes, where roll-like patterns are more localized and less stable, the azimuthal component of vorticity remains negative, but actual values are smaller.

Meridional flows self-develop in the simulations at $30^\mathrm{o}$ and $60^\mathrm{o}$ latitude. Northward meridional flows occupy most of the convection zone. The return flows occupy most of the bottom of the convection zone and the overshoot region (Fig.~\ref{fig:merid}). It is important to note a potential association of the meridional flows with the plasma ionization zones. In particular, at $60^\mathrm{o}$ latitude, simulations reveal a dip in the meridional flow profiles (red curves, Fig.~\ref{fig:merid}), which correspond to the bottom of the hydrogen and helium ionization zones. The potential effect of the He\,II ionization zone can also be noted in $30^\mathrm{o}$ and $60^\mathrm{o}$ latitudes for P$_{\rm rot}$=1~day. 

The presence of rotation impacts the thermodynamic properties of the convection zone (Table~\ref{tab:Prot-diff}). In particular, rotation suppresses temperature and density perturbations in the upper layers of the convection zone and causes an overall decrease of mean temperature in the whole convection zone. This effect increases for higher rotation rates and at higher latitudes (Fig.~\ref{fig:dTrot1-14}a,b). Considering temperature relative to the equator at the photosphere shows that the temperature deviates more strongly for slower rotation, as when temperature decreases by 220~K at $30^\mathrm{o}$ latitude and about 240~K at  $60^\mathrm{o}$ for a period of 14~days. In contrast, for faster rotation (P$_{\rm{rot}}=1$~day), the temperature decreases by about 90~K at $30^\mathrm{o}$ and about 202~K at $60^\mathrm{o}$ with a significantly stronger gradient between these latitudes. This temperature decrease at the photosphere associated with the rotation caused by the stellar radius decrease is well-known in astrophysics as gravitational darkening \citep{VonZeipel1924}.

Similarly, the temperature difference between $30^\mathrm{o}$ and $60^\mathrm{o}$ is stronger in the convection zone and the overshoot region. Below the photosphere, the temperature deviations from the equator are stronger for faster rotation. It is interesting to note that a small `bump' in temperature difference for faster rotation corresponds to the depths of the H and He~I ionization zone. The middle of the He~II ionization zone is manifested in the faster rotation case as a slight decrease in the temperature difference around 10~Mm depth. The temperature difference reaches a maximum near the bottom of the He~II ionization zone. A qualitatively similar behavior of the temperature difference is present at $60^\mathrm{o}$ latitude for slower rotation. 

In conclusion, we summarize the following common properties and effects for a moderate-mass main-sequence star with a shallow convection zone (about 28.5~Mm deep or about 2.81\% R$_{*}$) obtained with a series of models:
\begin{itemize}
    \item Stellar radius decreases by about 29~km at the equator and by about 58~km at higher latitudes for P$_{\rm{rot}}=1$~day.
    \item Development of differential rotation and meridional flows.
    \item The resulting rotation is slower than the imposed rotation rate in most of the convection zone (retrograde motions). It exceeds the imposed rotational velocity (prograde flows) in the bottom of the convection zone and the overshoot layer.
    \item The presence of strong differential rotation causes the formation of roll-like structures. The scale of roll-like structures and their lifetime decreases at higher latitudes.
    \item Meridional flows have a northward direction in most of the convection zone, and equatorward flows are at the bottom of the convection zone and in the overshoot layer.
    \item In the presence of rotation, the mean temperature is lower in the convection zone and the overshoot layer than in the non-rotating case. The most significant temperature deviations are in the photosphere. The temperature deviation gradually decreases with depth and agrees with the non-rotating case in the radiative zone.
    \item The simulation results show a temperature decrease for higher latitudes in the convection zone, known as gravitational darkening. At the upper layer of the overshoot region, the temperature increases with latitude and decreases again at the bottom of the overshoot region.
    \item The presence of rotation causes a slight decrease in the thickness of the He~I and H ionization zone and a shift of the He~II ionization zone closer to the photosphere.
\end{itemize}
In a follow-up paper, we will present a detailed analysis of changes in the dynamics and thermodynamic properties in the overshoot layer as well as other properties of the interface between the convection and the radiative zones, the so-called `tachocline.' 

\begin{acknowledgments}
 This work is supported by NASA Astrophysics Theory Program, Science DRIVE (Diversify, Realize, Integrate, Venture, Educate) Center Program (COFFIES Project``Consequences of Fields and Flows in the Interior and Exterior of the Sun"; 80NSSC22M0162). Resources supporting this work were provided by the NASA High-End Computing (HEC) Program through the NASA Advanced Supercomputing (NAS) Division at Ames Research Center.
\end{acknowledgments}


\end{document}